# An automated approach for consecutive tuning of quantum dot arrays

Hanwei Liu[1,2], Baochuan Wang[1,2,a)], Ning Wang[1,2], Zhonghai Sun[1,2], Huili Yin[1,2], Haiou Li[1,2,3], Gang Cao[1,2,3], and Guoping Guo[1,2,3,4]

[1]CAS Key Laboratory of Quantum Information, University of Science and Technology of China, Hefei 230026, China
[2]CAS Center for Excellence in Quantum Information and Quantum Physics, University of Science and Technology of China, Hefei 230026, China
[3]Hefei National Laboratory, University of Science and Technology of China, Hefei 230088, China
[4]Origin Quantum Computing Company, Hefei 230088, China
[a)]Author to whom correspondence should be addressed: bchwang@ustc.edu.cn

Recent progress has shown that the dramatically increased number of parameters has become a major issue in tuning of multi-quantum dot devices. The complicated interactions between quantum dots and gate electrodes cause the manual tuning process to no longer be efficient. Fortunately, machine learning techniques can automate and speed up the tuning of simple quantum dot systems. In this letter, we extend the techniques to tune multi-dot devices. We propose an automated approach that combines machine learning, virtual gates and a local-to-global method to realize the consecutive tuning of quantum dot arrays by dividing them into subsystems. After optimizing voltage configurations and establishing virtual gates to control each subsystem independently, a quantum dot array can be efficiently tuned to the few-electron regime with appropriate interdot tunnel coupling strength. Our experimental results show that this approach can consecutively tune quantum dot arrays into an appropriate voltage range without human intervention and possesses broad application prospects in large-scale quantum dot devices.

Semiconductor quantum dots (QDs) are promising candidates for quantum computation[1-3]. Spin qubits are constructed by controlling electrons (or holes) in quantum dot devices[4-9], which are controlled by tens of individual dynamic gate voltages with all of them being carefully set to isolate QDs to the few-electron regime[10-17]. Tuning quantum dot devices to the appropriate voltage configurations is the top priority, and even for a double-dot device, the tuning procedure is complicated, including the control of reservoirs, QDs and barriers. In addition, crosstalk between electrodes exacerbates the tuning complexity and reduces the characterization accuracy[18-23], which makes it difficult to tune a multi-dot system to the few-electron regime. Optimizing manipulation procedures to eliminate human intervention and crosstalk is now the key challenge in the optimal tuning of multi-dot devices.

In recent years, the idea of analyzing charge stability diagrams (CSDs) using convolutional neural networks (CNNs) has been proposed, based on which many experiments have achieved success in parameter extraction and device tuning, providing a solution for tuning quantum dot devices without human intervention[24-32]. This kind of data-driven machine learning (ML) technique can automate the tuning process through real-time analysis of the measurement results. There have been many proposals to accomplish this task, including using computer-supported algorithmic voltage control and feature matching for tuning[33-36], and some machine-learning-guide methods to reduce the number of measurements[37].

For large-scale quantum dot devices, as the number of QDs increases, it becomes harder to manually determine the appropriate gate-voltage configurations due to the complicated crosstalk between gate electrodes. To address this issue, we propose an automated approach that uses a CNN-based tuning process incorporated with optimization techniques to tune multi-dot devices. The CNNs are pretrained using the purposefully collected datasets to identify charge states from CSDs of DQDs. The optimization algorithms are designed for parametric fitting and are applied to different procedures. We extend the autotuning algorithm from a single DQD to a multi-dot array through an innovative local-to-global method, by dividing the array into multiple DQDs and using a sharing dot as a transition. Furthermore, virtual gates are introduced to help reduce crosstalk when tuning multiple quantum dots[38]. We develop a program based on this approach to tune a quadruple-quantum dot device, and the results exhibit good stability and accuracy, indicating its potential for application to large-scale quantum dot arrays.

Figure 1 shows a schematic of the autotuning process and a scanning electron micrograph of a quadruple-dot device identical to the one used in this experiment. This accumulation-mode device is fabricated with overlapping aluminum gate electrodes[39,40] electrically isolated from the Si/SiGe heterostructure by aluminum oxide. The device accommodates two single-electron transistors (SETs) acting as charge sensors[41] to detect the QD charge states using modulation signals. Opposing the two SETs, nine gate electrodes are fabricated to control the tunnel rates and electrochemical potentials of QDs. Plunger gates (labeled $P_i$, i = 1,2,3,4) accumulate electrons into QDs and shift the electrochemical potentials; barrier gates (labeled $B_i$, i = 1,2,3,4,5) separate QDs and adjust the tunnel rates between dots and to the reservoirs. The CSD shown in Fig. 1 is obtained by sweeping two plunger voltages while measuring the transconductance of the nearby charge sensor, and it contains information about the number of electrons in each dot and the interdot tunnel coupling strength between QDs.

To automate the tuning process and eliminate the need for human intervention, we choose to use the ML technique as an alternative. In particular, we pretrained two CNNs to identify charge states, where CNN1 is used for locating the few-electron regime, and CNN2 is used for determining the coupling between QDs. To optimize the autotuning process, it is necessary to map CSDs to charge states, so that the program can make corrections to the tuning process in time. Our goal is to find the appropriate gate-voltage configuration corresponding to the few-electron regime with medium coupling strength, under which the spin qubit system can be constructed. We first implement the autotuning algorithm in DQD, and then extend it to larger QD systems, combined with virtual gates and the local-to-global method. By dividing the multi-dot system into numbers of DQDs and applying virtual gates to ensure the independent

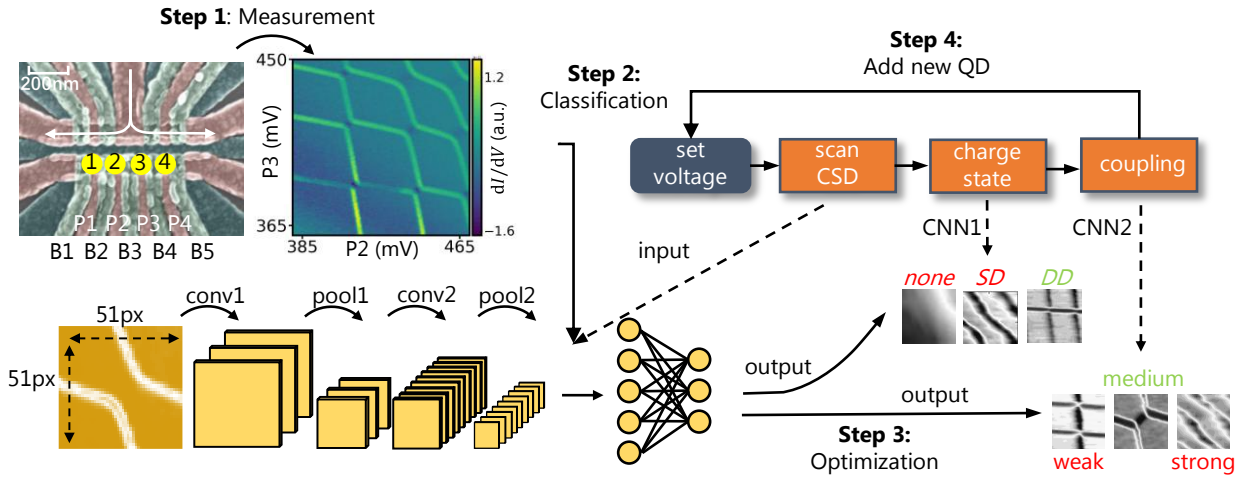

Fig. 1   Autotuning process schematic. The obtained CSDs are analyzed by two different CNNs to determine the charge states and inter-dot tunnel coupling, and only the CSDs with QDs being in the double-dot state and with medium coupling strength are reserved. After optimizing voltage configurations, new QDs are added, and a new cycle starts.

control of QDs, the consecutive tuning of the multi-dot system is successfully achieved.

The CNNs are built on AlexNet[42], and the training samples are cropped from a series of large-range CSDs. The training datasets consist of ~26,000 samples from existing experimental datasets mixed with ~3,000 samples from numerical simulations using the constant interaction model with different parameters. Sampled data from experiments contain noise, making CNNs more stable. Simulated data contribute to avoiding overfitting as they are clean and less distorted. All cropped samples are labeled as one of the three categories for charge state identification: *no dot* (*none*), *single dot* (*SD*) or *double dot* (*DD*). For coupling evaluation, the CNN is trained using the QD samples being in the double-dot state and the outputs are converted into a scalar corresponding to the coupling strength.

We define the output array $P_{charge} = [P_{none}, P_{SD}, P_{DD}]$ for charge state identification and the output scalar $P_{coupling} = 0 \cdot P_{weak} + 0.5 \cdot P_{medium} + 1 \cdot P_{strong}$ for coupling evaluation. Here, $P$ represents the probability given by CNN2. The loss function is defined by the cross-entropy of probabilities from assigned labels and CNN identifications, indicating the difference in probability distributions: $\text{loss} = -\sum_i P_{target}(i) \cdot \ln[P_{out}(i)]$. The subscript $i$ indicates the $i$th element of the probability array, $P_{target}$ is the assigned probability of the corresponding dataset given by humans and $P_{out}$ is the output probability to the same dataset given by the CNN. The loss function is minimized using the Adam optimizer[43] implemented in Python[44]. The schematic application of CNNs is illustrated in Fig. 1.

The CNN-based autotuning algorithm focuses on two key objectives: identifying few-electron regimes and tuning the coupling strength between QDs, which are related to voltages on the plunger and barrier gates respectively in our device architecture.

To make the autotuning algorithm efficient, we put restrictions on the voltage range within which the program can reach. The voltage range is expected to be set carefully: a small range contains little information while a large range slows the scan. In this experiment, the voltage range for each gate is determined by extracting the saturation and cutoff points from their pinch-off curves (see the supplementary material). In addition, we set a threshold for CNN identification to exclude abnormal results, making the tuning process less likely to be interrupted by external factors, especially noise. The size of the CSD obtained in the experiment is approximately $300 \text{ mV} \times 300 \text{ mV}$, and the diagram is analyzed by CNN1 to determine where few-electron regimes may exist. Note that the diagram spans a voltage range that is too wide, making it infeasible for the CNN to directly analyze the full diagram. We use a sliding window to traverse the full diagram and take a small piece out each time so that only a small subpart of the diagram is analyzed at a time.

Once the few-electron regime is confirmed, the program proceeds to evaluate the interdot tunnel coupling in the found few-electron regime. The quantified coupling given by CNN2 is normalized but has no physical meaning, the optimal value of which should be in the range of 0.3~0.7 (corresponding to the training samples containing anti-crossing regions where the interdot tunnel coupling strength is approximately 9-16 GHz). Values over 0.7 indicate over-strong coupling, and the program then decreases the voltage of the corresponding barrier gate and vice versa.

We select QD1 and QD2 from our quadruple-dot device to demonstrate the autotuning algorithm, and the results are illustrated in Fig. 2. The device is initialized manually, and voltage ranges and thresholds are set in advance. After confirming the presence of an electron transition, the program starts to search for the optimal voltage configuration that satisfies these two objectives.

In the stage of locating the few-electron regime, we calibrate a sliding window of size $45 \text{ mV} \times 45 \text{ mV}$ (21 px) to traverse the entire diagram with 8 mV (4 px) steps and CNN1 returns the possible charge state of each area (Fig. 2a and 2b). All the areas that are possibly in the double-dot state are selected as candidates so that the algorithm can select the one with minimum voltages as the few-electron regime. As shown in Fig. 2b, the positions where the charge transition lines intersect have higher probability values. For areas that contain only a single or several parallel transition lines or only the background noise, the corresponding returned probability values are lower than 70%, and the program only reserves areas with a probability higher than 80% (see the supplementary material).

Figures 2c and 2d show the process of tuning the interdot coupling of the found few-electron regime, and the red boxed area is used for identifying coupling strength. The change of barrier voltage shifts the charge transition lines on the CSD, and we dynamically adjust the sweep window with a change rate of $0.18\Delta V_B$ (the change rate depends on the device architecture) to cancel the shift of the anti-crossing. The coupling strength becomes stronger as the barrier gate voltage increases, and theoretically the output value given by CNN2 will also increase. Eight iterations are performed in the experiments and the output value gradually increases (Fig. 2d). The iteration stops due to the appearance of a qualified value of 0.37.

The autotuning success in the DQD system means it can be extended to large-scale QD systems. However, directly transplanting the autotuning algorithm to QD arrays is not feasible since the increasing number of QDs and gate electrodes creates more complex crosstalk, making it difficult for the tuned QDs to remain stable. To ensure that the tuning process does not affect the tuned QDs, a local-to-global method and a virtual gate technique are applied in our program. We demonstrate that the program using the CNN-based autotuning algorithm incorporated with the local-to-global method and virtual gates completes

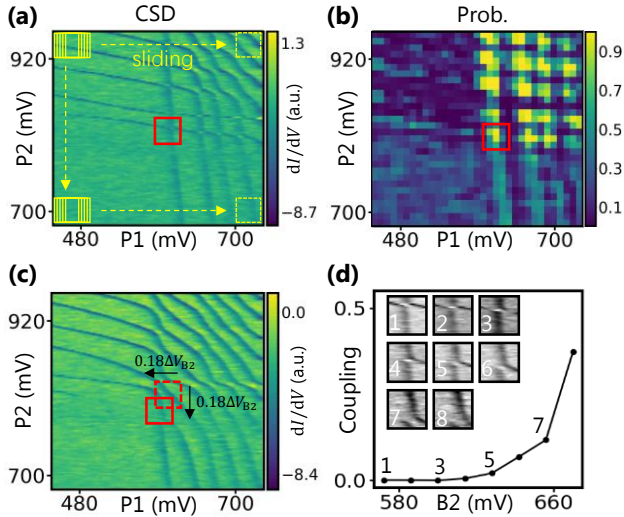

Fig. 2 DQD autotuning. (a) CSD of the initial state of QD1-QD2. A sliding window is used for traversing the diagram to identify the charge state. (b) Probabilities of QDs being in the double-dot state. The red box represents the found few-electron regime. (c) The CSD of the tuned QD1-QD2 system. The red box shows the sweep window for identifying the coupling value and its position changes with the barrier voltage. (d) Barrier tuning process. The CNN-identified coupling values increase as $V_{B2}$ increases. Insets: 8 CSDs of the found few-electron regimes when tuning Barrier B2.

consecutive tuning of a quadruple-dot device without any human intervention.

There are two main issues to consider in realizing automated tuning in multi-dot systems. One is to eliminate crosstalk between gate electrodes, and the other is to analyze CSDs obtained in high-dimensional gate-voltage spaces. To solve such problems, we propose the local-to-global method as a possible solution, the main idea of which is to divide multiple QDs into basic units, that is, the DQDs. Since the autotuning algorithm works well in tuning DQDs, the only concern is nondestructive tuning of the system after additional QDs are added.

The virtual gate technique can eliminate the crosstalk between gate electrodes and makes it possible to tune each QD independently, with which nondestructive tuning can be realized (Fig. 3a and 3b). The slopes of the charge transition lines are used to compute the cross-capacitance matrix $\boldsymbol{C}_\text{cross}$[38,45,46], with which the correspondence between virtual and physical gates can be established. The cross-capacitance causes physical gates to influence not only the electrochemical potential of corresponding QDs but also those of nearby QDs. A change in a virtual gate voltage corresponds to a linear combination of changes in several physical gate voltages, which can be described as

$$\begin{pmatrix} \Delta V_\text{U1} \\ \Delta V_\text{U2} \end{pmatrix} = \begin{pmatrix} 1 & c_{12} \\ c_{21} & 1 \end{pmatrix} \begin{pmatrix} \Delta V_\text{P1} \\ \Delta V_\text{P2} \end{pmatrix}, \tag{1}$$

where $\Delta V_\text{U}i$ and $\Delta V_\text{P}i$ are the virtual and physical gate voltages respectively, and the cross-capacitance matrix elements $c_{ij}$ can be extracted from the slopes of charge transition lines on CSDs. Note that the coupling between non-neighboring QDs is too weak to have an impact; therefore, tuning a series of DQDs formed by two neighboring QDs is sufficient.

Figure 3c shows a simple schematic of the local-to-global method. All tuned QDs are put into the idle state, that is, the plunger gate voltages are set to the few-electron regime, and barrier gates on both sides will be fixed in the tuned voltage in the subsequent tuning process. In experiments, it is only necessary to adjust voltages according to the electrode stacking order, which greatly simplifies the operation.

We demonstrate autotuning a linear quadruple-quantum dot array, and each pair of QDs in the array can be used as a fundamental tuning unit, to which the autotuning algorithm for DQDs can be applied. After tuning a DQD system, the virtual gates need to be updated based on the tuning results before adding a new QD. The cycle of "DQDs tuning – virtual gates updating – new dot adding" stops after the last QD

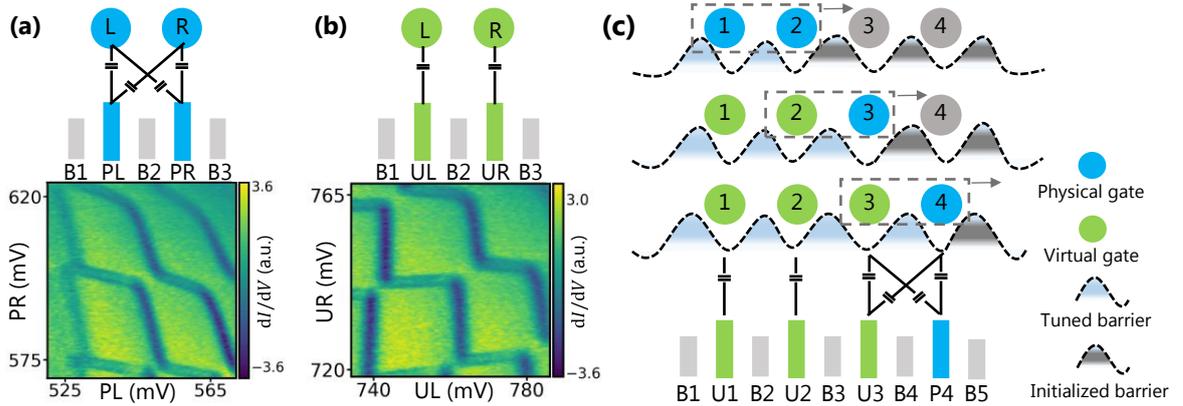

Fig. 3 Local-to-global method schematic representation. (a) and (b) show the comparison before and after applying virtual gates. (c) The local-to-global method. Gray dots represent QDs waiting to be tuned, blue dots are the DQDs being tuned, and green dots are QDs with virtual gates that have been established.

in the array has been tuned.

The QD array in this experiment is tuned in the order of QD1 to QD4, that is, starting with QD1 -

QD2, followed by QD2 - QD3 and QD3 - QD4. Two plunger gates and one barrier gate are involved to tune each QD pair. Note that the shared dot in neighboring pairs, QD2 in QD1 - QD2 and QD2 - QD3 for example, is paired with the added dot as a new DQD system. Virtual gates ensure that the tuned dots are not affected too much when adding new QDs.

In our cases, three DQDs are tuned by the program in sequence, making QD2 and QD3 the shared dots. Therefore, these two dots are tuned twice in two loops. To prevent conflicts between two different voltage configurations, all parameters are extracted from the newest results. This criterion relies on the fact that new QDs have much stronger effects on each other than tuned QDs since no virtual gates are applied to new dots. As a result, the QD1 parameters are determined in the first loop, QD2 in the second, QD3 and QD4 in the third.

Figure 4 shows the autotuning results of the quadruple quantum dot array and compares the charge states of few-electron regimes before and after adding new QDs. The boxed areas in Fig. 4a, 4b are the few-electron regimes, while the boxed area in 4c corresponds to the second-to-last transition line of QD4. The CSDs of few-electron regimes measured using virtual gates are shown in Fig. 4d, 4e and 4f. The QDs marked in blue and green are controlled by physical and virtual gates, respectively. Adding and tuning new QDs slightly influences the interdot tunnel coupling strength and shifts the charge transition lines of the tuned QDs, while the shape of the anti-crossing remains almost unchanged and cannot be identified by the CNN. Fig. 4g shows the corresponding identified coupling value during the tuning of barrier gate voltages.

The program efficiency is reflected in the time consumption. We evaluate the time usage in different stages of the entire autotuning process, and the scan time is excluded because it mainly depends on the measurement method. The CNN identification is efficient, and it takes milliseconds to return the identified results of each input diagram; that is, it needs approximately 5 seconds to traverse the entire diagram and obtain Fig. 2b that contains $30 \times 33 = 990$ data points. The number of iterations of tuning barrier gate voltages is determined by the initial state of a DQD system, and each iteration takes 5~10 seconds to change the barrier gate voltage and update the voltage range of the sweep window.

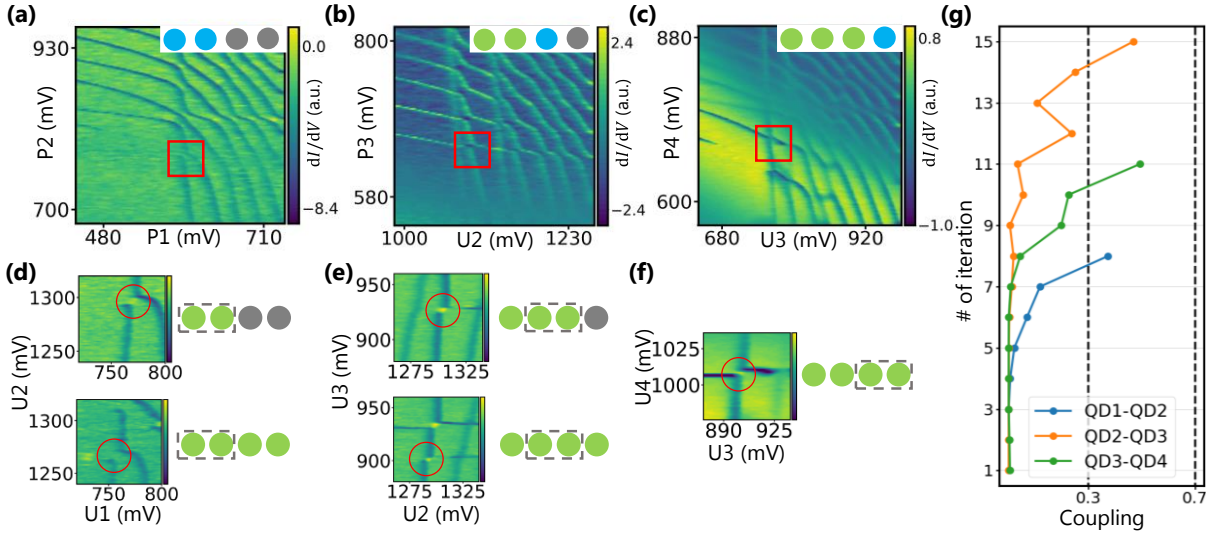

Fig. 4 Autotuning the quadruple-quantum dot array. (a) - (c) show the CSDs of the tuned QD1-QD2, QD2-QD3 and QD3-QD4 systems. (d) - (f) show the shift of the anti-crossing before and after the four QDs are tuned. QDs marked in blue are controlled by physical gates, while QDs marked in green are controlled by virtual gates. (g) The coupling evaluation changes in the iteration of tuning barrier voltages. The iteration stops when the CNN-evaluated coupling value is between 0.3 and 0.7.

Furthermore, the cross-capacitance matrix is updated by a computer using the Hough line transform[47] (see the supplementary material), which gives

$$C_{\text{cross}} = \begin{bmatrix} 1 & 0.211 & 0 & 0 \\ 0.484 & 1 & 0.307 & 0 \\ 0 & 0.153 & 1 & 0.198 \\ 0 & 0 & 0.472 & 1 \end{bmatrix}. \quad (2)$$

Every time a new QD is added, the matrix is updated to include the new dot. We simplify the matrix by ignoring barrier gates, so that only the elements corresponding to plunger gates need to be updated.

In conclusion, our experimental results verify that the automated approach using CNNs, virtual gates and the local-to-global method can consecutively autotune multi-dot systems. CNNs play an important role in identifying charge states, upon which the DQDs can be well-tuned by the autotuning algorithm. Furthermore, the local-to-global method helps extend the tuning capability to multi-dot systems. Despite the fact that the automated tuning approach cannot determine the exact inter-dot tunnel coupling strength, which is vital for high fidelity spin qubit gates and needs to be done through specific experiments such as photon-assisted-tunneling[48] (PAT), it still has great potential for eliminating human intervention when tuning large-scale quantum dot arrays.

## SUPPLEMENTARY MATERIAL

See the supplementary material for details: Figs. S1 and S5 shows the parameter extraction from pinch-off curves and CSDs, Figs. S2, S3 and S4 show the detailed results of the tuning process.

## ACKNOWLEDGMENTS

This work was supported by the Innovation Program for Quantum Science and Technology (Grant No. 2021ZD0302300), and the National Natural Science Foundation of China (Grant Nos. 12034018, 12074368, 92165207, and 61922074). This work was partially carried out at the University of Science and Technology of China Center for Micro and Nanoscale Research and Fabrication.